\documentclass[12pt]{article}
\usepackage{pic03}
\usepackage{graphicx}

\begin{document}

\title{\bf The Muon Decay at the Two Loop Level of the Electroweak
  Interactions.}
\author{
M. Czakon \\
{\em Department of Field Theory and Particle Physics,} \\
  {\em Institute of Physics, University of Silesia, Uniwersytecka 4,} \\
  {\em PL-40007 Katowice, Poland}}
\maketitle

\baselineskip=14.5pt
\begin{abstract}
  The progress in the evaluation of the muon decay lifetime is
  reviewed. The electroweak bosonic corrections are given together with
  the respective shift of the $W$ boson mass. After inclusion of the
  recently corrected fermionic contributions, this requires an updated
  fitting formula for the prediction of $M_W$.\footnote{The results
    presented in this contribution were obtained in collaboration with
    M. Awramik.}
\end{abstract}

\baselineskip=17pt

\section{Introduction}

The muon lifetime is one of the key observables of today's particle
physics. Not only is it measured very precisely, since the current
experimental error is 18 ppm, but can be described to competing
accuracy within the Standard Model, giving rise to a strong
correlation between the masses of the heavy gauge bosons. As a low
energy process, the decay is expected to be governed by an effective
interaction involving only the electron, muon and their respective
neutrinos. The dynamics of the system should be corrected mostly by
QED, whereas the electroweak interactions determine solely the size of
the coupling constant.

The calculation at the two loop level has been completed recently
with the evaluation of the electroweak bosonic corrections
\cite{Awramik:2002wn}. Subsequently, the fermionic part, which was
first obtained in \cite{Freitas:2000gg} has been recalculated
\cite{Awramik:0000}. A difference between the two results has been
found and the original result \cite{Freitas:2000gg} corrected. There
is now agreement on all parts of the two loop contributions.
\begin{figure}[h!]
  \centerline{
    \includegraphics[width=7cm,angle=270]{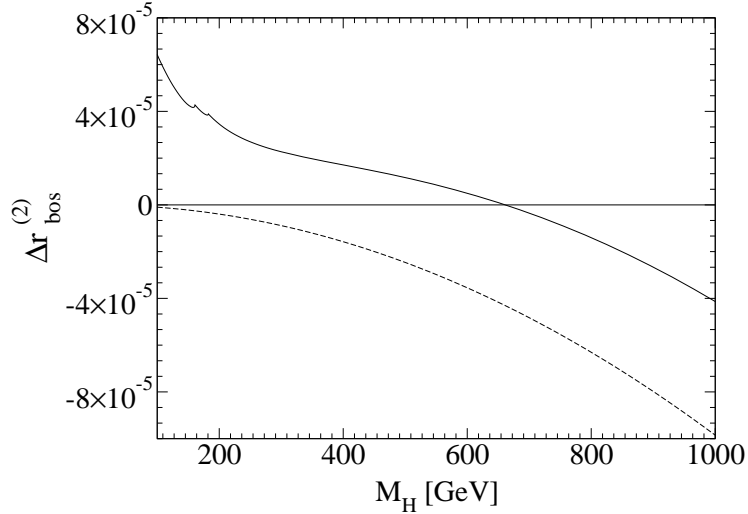}}
  \caption{\it The complete bosonic corrections (solid line) together with the
    leading term (dashed line) in the large Higgs boson mass
    expansion. \label{result}}
\end{figure}

\section{Two loop contributions}

The main complication in the calculation of the bosonic contributions
is connected with the separation of the QED corrections to the Fermi
model. This problem has been solved with the use of the effective
theory language, where the full lagrangian 
\begin{equation}
  {\cal L}_{\rm SM} = {\cal L}_{\rm SM}(\alpha, \alpha_s,
  m_{\nu}, m_l, m_q, M_W, M_Z, M_H, m_t, V_{\rm MSN}, V_{\rm CKM}),
\end{equation}
has been replaced with the Fermi lagrangian
\begin{eqnarray}
  \label{Leff}
        {\mathcal L}_{\rm eff} &=& {\mathcal L}_{\rm kin}(\nu)+
        {\mathcal L}_{\rm QED}(\alpha^0,m_l^0,m_q^0,l^0,q^0,A^0_\mu)
        +{\mathcal L}_{\rm QCD}(\alpha_s^0,m_q^0,q^0,A^{a,0}_\mu)
        \\ \nonumber
        &+& \frac{G_F}{\sqrt{2}}\;\;\overline{e^0}
        \gamma^\alpha (1-\gamma_5) \mu^0 \times \overline{\nu_\mu}
        \gamma_\alpha (1-\gamma_5) \nu_e,
\end{eqnarray}
and the amputated renormalized Green functions in both theories have
been required to be equal up to higher order terms in an expansion in
the $W$ boson mass
\begin{equation}
  {\mathcal G}_{\rm SM} = {\mathcal G}_{\rm eff}
  +{\mathcal O}(1/M_W^4).
\end{equation}
The above {\em matching equation} has the advantage that one can take
any suitable masses and momenta of the light particles and
subsequently expand in them. For the calculation, all of these
parameters were put to zero and no expansion was necessary. This lead
to problems with the Dirac gamma matrix structure in the four-fermion
operators, which were solved by a suitable projection
operator. However, the procedure has also been tested with fermion
masses as infrared regulators and agreement was found.

\begin{figure}[h!]
  \centerline{
    \includegraphics[width=11cm]{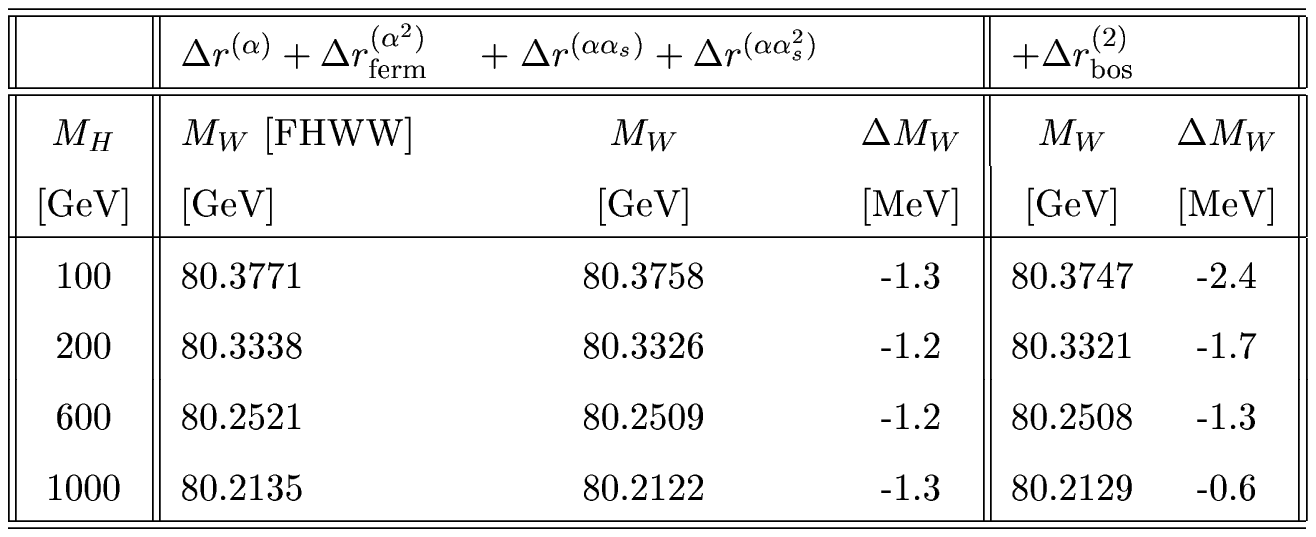}}
  \caption{\it The impact of the bosonic and the corrected
    fermionic contributions on the prediction of the $W$ boson mass. The
    second column gives the result from \cite{Freitas:2000gg}.
    \label{impact}}
\end{figure}

The result for the pure bosonic contributions is given in
Fig.~\ref{result}. The impact of this class on the prediction of the
$W$ boson mass as well as that of the fermionic part is summarized in
Fig.~\ref{impact}.
\section{Acknowledgments}
This work was supported in part by DESY Zeuthen and the KBN Grant 2P03B01025.

\end{document}